\documentclass[11pt,a4paper,useAMS,usenatbib]{emulateapj}
\usepackage{epsfig}
\usepackage{amsmath}
\usepackage{natbib}

\begin{document}

\title{Connecting dark matter halos with \\  the galaxy center and the supermassive black hole}

\author{\'Akos Bogd\'an\altaffilmark{1} and Andy D. Goulding\altaffilmark{1,2}}
\affil{\altaffilmark{1}Harvard Smithsonian Center for Astrophysics, 60 Garden Street, Cambridge, MA 02138, USA; abogdan@cfa.harvard.edu}
\affil{\altaffilmark{2}Department of Astrophysical Sciences, Princeton University, Princeton, NJ 08544, USA}
\shorttitle{LINKING HALOS TO THEIR ELLIPTICAL GALAXIES}
\shortauthors{BOGD\'AN \& GOULDING}

\begin{abstract}
Observational studies of nearby galaxies have demonstrated correlations between the mass of the central supermassive black holes (BHs) and properties of the host galaxies, notably the stellar bulge mass or central stellar velocity dispersion. Motivated by these correlations, the theoretical paradigm has emerged, in which BHs and bulges co-evolve. However, this picture was challenged by observational and theoretical studies, which hinted that the fundamental connection may be between BHs and dark matter halos, and not necessarily with their host galaxies. Based on a study of 3130 elliptical galaxies -- selected from the Sloan Digital and \textit{ROSAT} All Sky Surveys -- we demonstrate that the central stellar velocity dispersion exhibits a significantly tighter correlation with the total gravitating mass, traced by the X-ray luminosity of the hot gas, than with the stellar mass. This hints that the central stellar velocity dispersion, and hence the central gravitational potential, may be the fundamental property of elliptical galaxies that is most tightly connected to the larger-scale dark matter halo. Furthermore, using the central stellar velocity dispersion as a surrogate for the BH mass, we find that in elliptical galaxies the inferred BH mass and inferred total gravitating mass within the virial radius (or within five effective radii) can be expressed as $M_{\rm{BH}} \propto M_{\rm tot}^{1.6^{+0.6}_{-0.4}} $ (or $M_{\rm{BH}} \propto M_{\rm{5r_{eff}}}^{1.8^{+0.7}_{-0.6}}$). These results are consistent with a picture in which the BH mass is directly set by the central stellar velocity dispersion, which, in turn, is determined by the total gravitating mass of the system.
\end{abstract}


\section{Introduction}
Supermassive black holes (BHs), located at the center of every massive galaxy, are believed to have a profound effect on the evolution of their host galaxies.  Rapidly growing BHs shine as active galactic nuclei (AGNs), and can release copious amounts of energy into their surroundings, which can heat and expel their gas supply, thereby quenching star formation and the BH growth \citep{silk98,king03,wyithe03,hopkins06}. Energetic feedback from BHs was suggested to play a pivotal role in building the present-day observed galaxy luminosity function \citep{croton06,faber07}, and hence BHs are a key component of modern galaxy formation theories. 

Observational studies of nearby galaxies have pointed out the existence of scaling relations between the BH mass ($M_{\rm{BH}}$) and various properties of the host galaxies, in particular with the stellar bulge mass, $M_{\rm{bulge}}$ \citep{magorrian98,haring04}; or central stellar velocity dispersion, $\sigma_{\rm{c}}$ \citep{gebhardt00,ferrarese00,gultekin09}. To explain the existence of these relatively tight scaling relations, a widely accepted theoretical paradigm has been established, in which BHs grow in a self-regulated manner and co-evolve with their host bulges from early epochs to the present day \citep{silk98,fabian99,king03,hopkins06}. However, several observational and theoretical studies revealed correlations between the BH mass and the total gravitating mass, thereby hinting that dark matter halos may govern the BH growth \citep{ferrarese02,bandara09,booth10,booth11,bogdan12}. Follow-up studies demonstrated that in late-type or bulgeless galaxies the dark matter halo and the BH mass do not correlate tightly \citep{kormendy11,sun13}. The absence of a tight relation in these galaxies may be due to low-mass or disk-dominated galaxies being less efficient at growing BHs, and dynamical effects that increase the scatter in the $M_{\rm{BH}}-M_{\rm{tot}}$ relation for low-mass systems \citep{volonteri11}. However, observationally it is not well established, whether dark matter halos play a major role in driving the BH growth in elliptical galaxies. Therefore, to fully explore the connection between the BH mass and total halo mass, it is essential to perform a dedicated study focusing on elliptical galaxies.
 
The main difficulty in conclusively determining whether the growth of BHs in elliptical galaxies is driven by stellar bulges or their large-scale dark matter halos arises from the demanding nature of BH and halo mass measurements. Specifically, dynamical BH mass measurements have only been obtained for $\sim80$ galaxies, with about half of these BHs being located in elliptical galaxies \citep{gultekin09,mcconnell13}. Of these systems, only a handful have precisely measured halo masses \citep{humphrey06}, which is insufficient to probe the importance of dark matter halos. Additionally, the local sample of galaxies with available BH and halo mass measurements does not offer a representative sample because galaxies residing in clusters are overrepresented and include several brightest cluster galaxies.  Therefore, instead of using individual galaxies, we may opt to study the {\it average} properties of a large elliptical galaxy sample.

In this paper, we probe, based on a statistically significant galaxy sample, whether BHs located in elliptical galaxies exhibit a correlation with their total gravitating mass. Because the direct determination of the BH mass ($M_{\rm BH}$) and halo mass is not feasible for a large galaxy sample, we rely on proxies. For massive BHs, both observational and theoretical considerations suggest that the central stellar velocity dispersion provides a good measure of the BH mass \citep{gultekin09,mcconnell13}; for details see Section \ref{sec:bhproxy}. In the framework of the Sloan Digital Sky Survey (SDSS), a large number of nearby galaxies have been explored in a uniform manner, and SDSS observations provide the required accurate velocity dispersion measurements. For total halo mass measurements, the X-ray luminosity of the hot gas can be used as a robust tracer \citep{mathews06,kim13}; for details see Section \ref{sec:dmproxy}. To trace the hot gas component of ellipticals, we employ data from the \textit{ROSAT} All Sky Survey (RASS). However, because of the relatively short individual RASS observations, most passive elliptical galaxies beyond $z\sim0.01$ remain undetected. Therefore, the X-ray properties of individual galaxies cannot be explored with RASS. However, when using the position information of a sufficiently large sample of SDSS galaxies, and stacking the {\it individual} X-ray observations, the resulting {\it average} galaxy population may show significant X-ray emission. Thus, to measure the X-ray luminosity, and hence the total halo mass of SDSS galaxies, we perform a stacking analysis. A stacking analysis has the additional advantage of probing the {\it average} properties of the galaxy population. Such a study naturally removes the distorting effects of outliers, as well as guards against biased selection techniques that are inherent in many studies focusing on small numbers of individual objects.

The paper is structured as follows. In Section 2 we introduce and describe the sample of elliptical galaxies. In Section 3 we present the analysis of the \textit{ROSAT} data, and in Section 4 we present our main result and show that the central stellar velocity dispersion of elliptical galaxies exhibits a very tight correlation with the X-ray luminosity of the hot gaseous component. In Section 5 we discuss that the central velocity dispersion and the gas X-ray luminosity can robustly trace the BH and {total halo masses}, and hence our results also imply that the BH mass tightly correlates with the total halo mass. We conclude in Section 6. 

\begin{table*}
\caption{Observed Galaxy Properties in Each Bin.}
\begin{minipage}{18.5cm}
\renewcommand{\arraystretch}{1.3}
\centering
\begin{tabular}{c c c c c c c c c}
\hline 
$ M_{\rm{\star}} $& $M_{\rm{\star}} $ & $\sigma_{\rm c}$  & Galaxy &   Distance& $N_{H}$ &   Exposure  & $L_{\rm X,LMXB} $ &  $L_{\rm X,gas} $       \\ 
($10^{10} \ \rm{M_{\odot}}$)  &  ($ 10^{10} \ \rm{M_{\odot}}$)  &  ($\rm{km \ s^{-1}}$)  &  \#  &  (Mpc)   & ($10^{20} \ \rm{cm^{-2}} $) &  (ks)  & ($10^{40} \ \ \rm{erg \ s^{-1}}$) & ($10^{40} \ \rm{erg \ s^{-1}}$)   \\ 
  (1)    &   (2)    &            (3)           &   (4)      &     (5)     &      (6)                 &   (7) & (8)   &(9) \\
\hline 
$ 1.00-2.51 $&1.58  &  78.7  & 265 &   $ 168.9  $ &   2.03 &  117.2 &    $0.04$ &  $ -  $\\
$ 1.00-2.51 $& 1.70  & 111.8  & 501 &   $ 165.4  $ &    2.04 &  212.2 &    $0.04$ & $ -  $\\
$ 1.00-2.51$ & 1.62  &  145.5 & 192 &   $ 169.1  $ &    2.05 &  85.8  &    $0.04$ & $ - $\\
$ 2.51-6.31 $& 3.89  & 115.2  & 246 &   $ 179.5  $ &   2.07 &  105.2 &    $0.09$ & $<1.0$\\
$ 2.51-6.31 $& 3.98  & 149.1  & 717 &   $ 172.0  $ &   2.06 &  301.1 &    $0.09$ & $ 1.2\pm0.8$\\
$ 2.51-6.31$ & 3.80  & 193.5  & 321 &   $ 164.7  $ &   2.03 &  142.8 &    $0.09$ & $ 4.2\pm1.1$ \\
$ 6.31-15.85 $& 9.12  & 155.6  & 102 &   $ 176.7  $ &    2.08 &  40.9  &    $0.20$ & $ 2.6\pm2.2$ \\
$ 6.31-15.85$& 9.12  &  197.3 & 567 &   $  172.6 $ &   2.04 &  239.7 &    $0.22$ & $ 6.4\pm0.9$ \\
$ 6.31-15.85$ & 8.70  & 254.0  & 117 &   $ 161.9  $ &    2.04 &  48.2  &    $0.21$ & $ 12\pm2$ \\
$ 15.85-39.81 $& 19.50  &  249.6 & 102 &   $ 169.8  $ &   2.07 &  43.7  &    $0.47$ & $ 14\pm2$ \\ 
\hline \\
\end{tabular} 
\end{minipage}
\textit{Note.} Columns are as follows. (1) Stellar mass range of galaxies  in units of $10^{10} \ \rm{M_{\odot}}$. (2) Median stellar mass  in units of $10^{10} \ \rm{M_{\odot}}$. Note that for ellipticals the stellar mass is equivalent with the stellar bulge mass. (3) Median central stellar velocity dispersion given by SDSS. (4)  Number of galaxies in each bin. (5) Median distance. (6) Median line-of-sight column density derived from the HI map provided by the LAB survey \citep{kalberla05}.  (7) Combined exposure time of the galaxies in each bin. (8) X-ray luminosity of the population of unresolved LMXBs in units of $10^{40} \ \rm{erg \ s^{-1}}$. Their luminosity is derived from the average LMXB luminosity function \citep{gilfanov04}.  (9) Observed X-ray luminosity of the hot gas  in units of $10^{40} \ \rm{erg \ s^{-1}}$. The contribution of LMXBs from Column (8) is subtracted. Note that $L_{\rm X,gas} $ for the first three rows are consistent with $L_{\rm X,gas} =0$.  \\

\label{tab:list1}
\end{table*}

\section{Sample selection}
To construct a statistically significant galaxy sample, we rely on the SDSS Data Release 10 and the RASS data. First, we selected all SDSS galaxies that are within the redshift range of $z=0.01-0.05$, have a stellar mass measurement, and have a stellar velocity dispersion measurement using the MPA-JHU analysis of the SDSS DR10 catalog \citep{kauffmann04,brinchmann04}. The lower redshift boundary was introduced to exclude the most nearby galaxies, which given their proximity appear as the brightest on the sky (see Section \ref{sec:rosat}). The higher redshift boundary is necessary to ensure the reliability of the central velocity dispersion measurements. Specifically, at $z>0.05$ the CaII triplet feature (8498\AA; 8542\AA; 8662\AA) is redshifted out of the observable range of SDSS spectroscopy (3900-9100\AA), thereby leading to larger uncertainties in the velocity dispersion measurement for galaxies beyond $z>0.05$.

In this work, we focus on the population of elliptical galaxies. A particular advantage of studying ellipticals is that for these systems the  stellar mass is equivalent to the stellar bulge mass. Therefore, our results are not affected by uncertainties associated with bulge-disk decomposition, such as with studies of late-type galaxies. To study elliptical galaxies in the SDSS catalog, it is essential to robustly determine the morphology of the sample galaxies.

Traditionally, elliptical galaxies were identified based on their location in the galaxy color--magnitude diagram \citep{bell04}. However, recent results demonstrate that both early-type and late-type galaxies are distributed across the entire color range \citep{schawinski14}, making it difficult to identify early-type galaxies solely using their galaxy colors. Therefore, the morphology of the sample galaxies must be determined by visual inspection on a case-by-case basis. To carry out this task for our large galaxy sample, we relied on the data provided by the Galaxy Zoo project \citep{lintott08}. In the framework of Galaxy Zoo, citizen scientists visually inspected the morphology of a vast number of SDSS galaxies, allowing us to unbiasedly select elliptical galaxies in a galaxy-color-independent manner. To ensure that our sample only contains ellipticals, we included those systems that were deemed to be as elliptical galaxies by at least $70\%$ of the citizen scientists. This definition resulted in a parent sample of 4592 elliptical galaxies that are observed with SDSS and are within the redshift range of $z=0.01-0.05$. Additionally, we visually inspected all 4592 galaxies, and Ð in agreement with the Galaxy Zoo project Ð found that all selected galaxies exhibit an elliptical morphology. As a caveat, we note that the sample may contain S0 galaxies in projection, which are difficult to remove robustly. Moreover, we established that none of the galaxies in our parent sample appears to be substantially star forming.

According to previous observational studies, low-stellar-mass systems cannot retain a significant amount of hot X-ray gas in their gravitational potential well  \citep{osullivan01,bogdan11}. Therefore, we excluded all galaxies with stellar masses below $\log M_{\rm{\star}}/\rm{M_{\odot}}=10.0 $. 

To ensure that any observed X-ray emission within the stacked data is not being produced by AGNs, we used two methods to identify and remove AGNs. First, we applied typical optical emission line diagnostic diagrams from the SDSS optical spectroscopy. Here we choose to use both the [N{\sc ii}]/H$\alpha$--[O{\sc iii}]/H$\beta$ and [O{\sc i}]/H$\alpha$--[O{\sc iii}]/H$\beta$ diagnostics \citep{kewley06,schawinski07}. When combined, these ionization ratio diagrams efficiently and cleanly separate AGN emission from lower-ionization emission, such as that from star formation. Those sources that lie above the relevant demarcation curves are identified as AGNs and removed. Second, all galaxies that were directly detected in the RASS images were further removed from the sample. However, as evidenced by inspection of the SDSS optical images, these particular sources were the brightest cluster galaxies at (or close to) the center of large groups and clusters. Based on these methods, we identified and removed 527 AGNs from the sample.

In addition to these, we searched through the SDSS and RASS images and filtered 299 sources that were contaminated by foreground or background objects, such as a nearby galaxy in projection or a luminous background AGN. Such sources could contaminate the stacks and hence must be removed. We also searched for galaxies that had a companion with similar redshift within a $3\arcmin$ projected distance. The dark matter halos of these interacting galaxies may not be virialized, and hence the observed X-ray luminosity cannot be robustly used to trace their halo mass. We identified 185 interacting sources in our sample. Finally, we removed those 451 galaxies that were located in rich groups or clusters. The detected X-ray emission in these galaxies would likely be dominated by the large-scale hot diffuse group/cluster gas and not by hot gaseous coronae of the sample galaxies. The final sample contains 3130 elliptical galaxies.

\section{Analysis of the \textit{ROSAT} data}
\label{sec:rosat}
The RASS explored the entire sky in the soft X-ray regime, providing the only all-sky survey at X-ray energies to date. In the framework of the RASS, 1378 fields were observed, the majority performed in scanning mode, which resulted in fields with $6.4\degr \times 6.4\degr$ length sides. The average exposure time of the RASS observations is approximately $430$ s, and hence most non-AGN galaxies beyond $z\sim0.01$ are not detected. However, despite the shallow average exposure times of the RASS observations, the most nearby galaxies may be detected. These individual detections could bias the stacked signal, and hence they were excluded from our study. Because the individual observations may have significantly shorter or longer exposures than the average value, it is necessary to account for the variations in the exposure times. To this end, we utilized the exposure maps that are provided with each observation. The combined exposure time of our 3130 sample galaxies amounts  to $\sim1.3$ Msec.  

Although \textit{ROSAT} collected photons across the $0.1-2.4$ keV energy range, in our analysis we opted to use the $0.4-2.4$ keV band for the following three reasons. First, the \textit{ROSAT} effective area sharply decreases below $0.4$ keV, resulting in fewer detected source photons at lower energies.  Second, the instrumental background level is relatively high in the $0.1-0.4$ keV band, so the signal-to-noise ratio may be reduced when the softest energy range is also used. Third, given the average level of Galactic absorption ($N_{\rm{H}} = 2.1\times10^{20} \ \rm{cm^{-2}}$) for our galaxy sample, we do not expect significant emission from the hot gaseous component below $0.4$ keV energy.  

Compared to present-day X-ray telescopes, the point-spread function of \textit{ROSAT} is relatively broad, which must be considered when choosing the source and background extraction regions. To this end, we used circular regions with a four-pixel ($180 \arcsec$) radius centered on the stacked images to extract the source region. This region confines about $83\%$ of the source counts \citep{boese00,bohringer14}. This region is optimal because it contains a large fraction of the source counts, while it also minimizes the number of background counts. We note that choosing a somewhat smaller or larger source extraction region does not affect our results in any significant way. To account for the source counts falling outside of our extraction region, we increased the observed net count rates by $1/0.83=1.20$. We subtract the background components by using local background regions. Utilizing local background regions is advantageous because both the instrumental and sky background components can be precisely subtracted. Therefore, we used circular annuli with a $8-14$ pixel ($360\arcsec-630\arcsec$) radii to account for the background components. Within these regions we do not expect a notable contribution from the stacked signal, thereby ensuring a precise background subtraction.

\begin{figure}[!]
  \begin{center}
    \leavevmode
      \epsfxsize=8.7cm\epsfbox{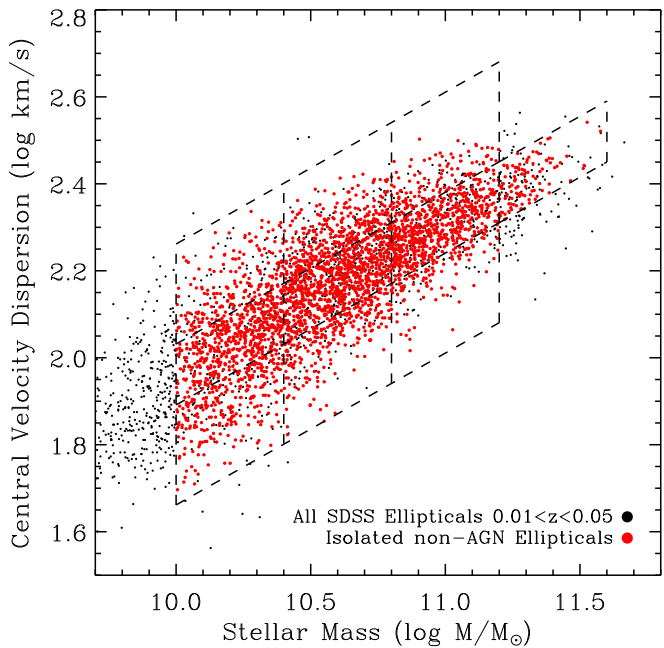}
      \caption{Analyzed sample of elliptical galaxies. Black dots show the velocity dispersion as a function of the stellar mass for the parent sample of 4592 elliptical galaxies observed by SDSS. The red dots depict the 3130 galaxies selected in our stacking analysis after removal of all interacting systems, galaxies located in rich groups or clusters, images contaminated by foreground or background objects, and AGN detections. The dashed regions show the selection regions as defined by the stellar mass and central velocity dispersion.}
     \label{fig:fig2}
  \end{center}
\end{figure}

To perform our stacking analysis, we produced $24\arcmin \times 24\arcmin$ cutout images of the individual X-ray images and  the corresponding exposure maps. The cutout X-ray images (and their respective exposure maps) were summed to produce stacked X-ray images of the galaxies in our sample for a given stellar mass and central stellar velocity dispersion (see Section \ref{sec:bins}).

\section{Results}
\subsection{Binning the Galaxy Sample}
\label{sec:bins}
Our large galaxy sample allows us to probe whether the central stellar velocity dispersion and the X-ray luminosity exhibit a tight correlation with each other. To this end, we grouped the galaxy sample based on their stellar mass and central velocity dispersion. We split the sample into four equal-width stellar mass bins in the $\log M_{\rm{\star}}/\rm{M_{\odot}}=10.0-11.6$ stellar mass range. The galaxies within each of these bins are further subdivided according to their velocity dispersion (Figure \ref{fig:fig2}). Specifically, we grouped $68\%$ of galaxies that follow the average stellar velocity dispersion--stellar mass relation in the central region of each mass bin, while galaxies with seemingly low or high velocity dispersions, relative to their stellar mass, are located below or above the central regions, respectively. Because of the low number of galaxies in the $\log M_{\rm{\star}}/\rm{M_{\odot}}=11.2-11.6$  stellar mass range, we only defined one group. The average properties of the galaxies residing in each bin are tabulated in Table \ref{tab:list1}. 

\begin{figure}[!]
  \begin{center}
    \leavevmode
      \epsfxsize=8.7cm\epsfbox{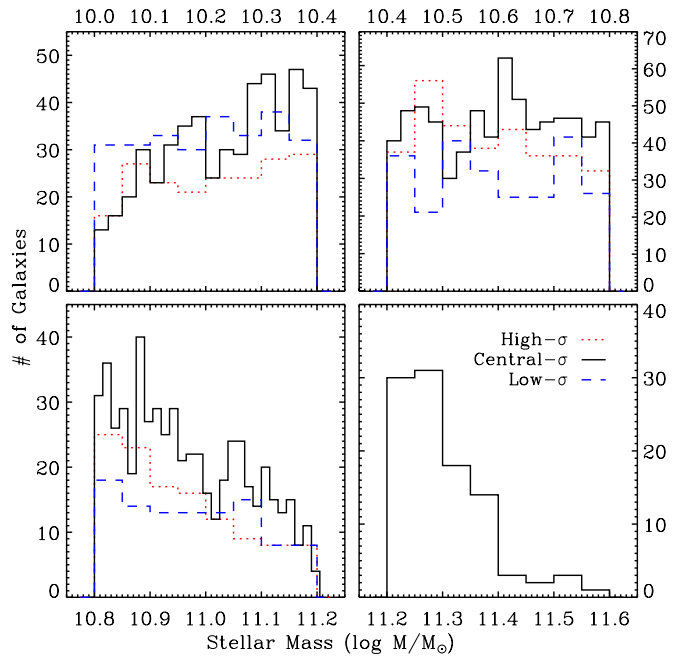}
      \caption{Four panels show the stellar mass distributions for each galaxy bin. Within each stellar mass bin the mass distributions are similar, independent of the median velocity dispersions. Based on Kolmogorov-Smirnov tests, we find no statistically significant evidence that any of the stellar velocity dispersion samples are drawn from different distributions. Note that in the $\log M_{\rm{\star}}/\rm{M_{\odot}}=11.2-11.6 $ stellar mass range, only one group is defined because of the low number of galaxies.}
     \label{fig:fig1}
  \end{center}
\end{figure}

\begin{figure*}[!t]
  \begin{center}
    \leavevmode
      \epsfxsize=4in\epsfbox{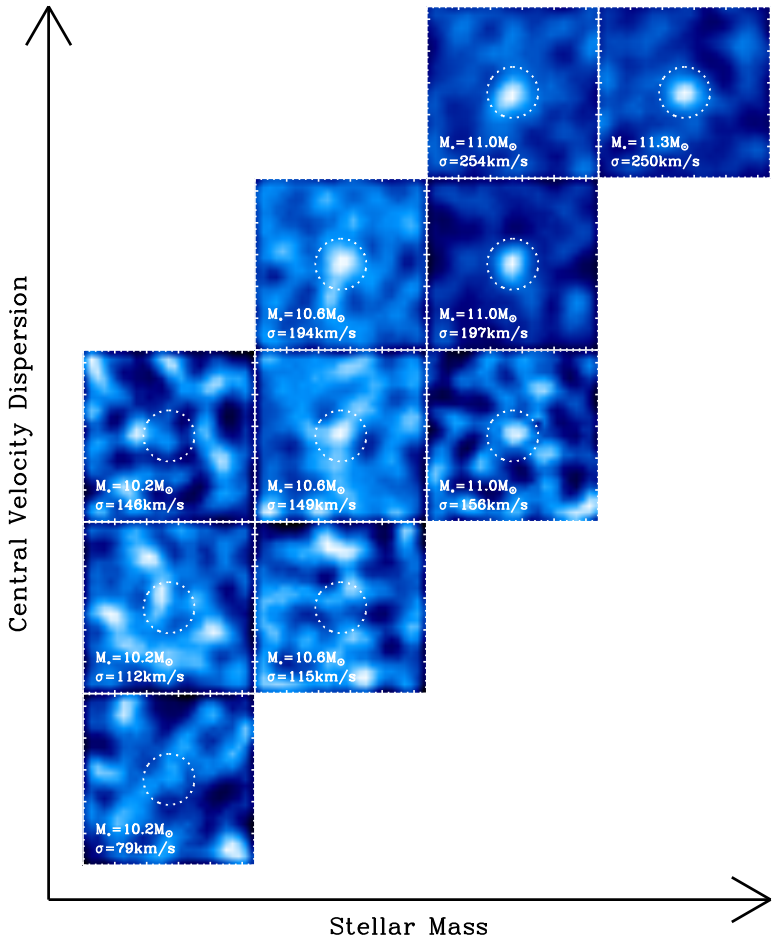}
      \caption{Schematic diagram of stacked RASS X-ray images for the 10 bins defined by their stellar mass and velocity dispersions. No exposure correction is applied. The images were smoothed with a Gaussian kernel of three-pixel size. The circular region has a radius of four pixels and represents the source extraction region. The bins with high ($>190 \ \rm{km \ s^{-1}}$) median velocity dispersion exhibit a statistically significant, $4.2\sigma-7.3\sigma$, detection, and bins with low ($<190 \ \rm{km \ s^{-1}}$) median velocity dispersion either show a moderate detection ($\sim2\sigma$) or a nondetection. Note that in the lowest stellar mass bin ($\log M_{\rm{\star}}/\rm{M_{\odot}}=10.0-10.4$), no statistically significant source is detected, which is partly due to the shallow potential well of these galaxies, and partly due to the low effective depth of the stacked data in the bin characterized with the highest ($149 \ \rm{km \ s^{-1}}$) velocity dispersion.}
\vspace{0.5cm}
     \label{fig:fig3}
  \end{center}
\end{figure*}

Given that we use the stellar mass as a major criterion to divide the galaxy sample into multiple bins, we investigate the stellar mass distributions within each bin. The obtained stellar mass distributions are shown in Figure \ref{fig:fig1}, which demonstrates that within the same stellar mass interval, the stellar mass distributions are nearly identical. To statistically probe if the galaxy samples are drawn from the same parent population within each stellar mass bin, we performed a series of Kolmogorov-Smirnov tests. We obtained probabilities $>0.09$ for the comparison of each low-, central-, and high-velocity dispersion samples within the same stellar mass bins. This implies that we have no statistically significant evidence to suggest that any of the velocity dispersion samples are drawn from separate distributions. 

The average redshift of the galaxies in our sample is $z\approx0.0415$, which corresponds to a distance of $170.1$ Mpc. To ensure that possible differences in the redshift distribution are not affecting our results, we investigated  the distance distributions of each galaxy bin. In particular, we performed Kolmogorov-Smirnov tests, which pointed out that there is no evidence at the $>1\%$ level that the $\log M_{\rm{\star}}/\rm{M_{\odot}}=10.0-10.4$ and $\log M_{\rm{\star}}/\rm{M_{\odot}}=10.4-10.8$  and $\log M_{\rm{\star}}/\rm{M_{\odot}}=10.4-10.8$  and $\log M_{\rm{\star}}/\rm{M_{\odot}}=10.8-11.2$  mass bins are not drawn from the same distributions. However, statistically we would reject $H_{\rm 0}$ at the $1\%$ level for the comparison between the $\log M_{\rm{\star}}/\rm{M_{\odot}}=10.0-10.4$  and $\log M_{\rm{\star}}/\rm{M_{\odot}}=10.8-11.2$ samples because of  a slightly larger population of more massive systems at higher redshift, which is to be  expected for any flux-limited sample. But we would still accept $H_{\rm 0}$ at the $0.1\%$ level for these bins, so any difference can still be considered marginal. The median distances of the galaxies in each bin are close to $170.1$ Mpc, and none of the bins deviates more than $\sim5\%$ from this average value (Table \ref{tab:list1}).
  
\begin{figure*}[!t]
  \begin{center}
\includegraphics[width=6.5in]{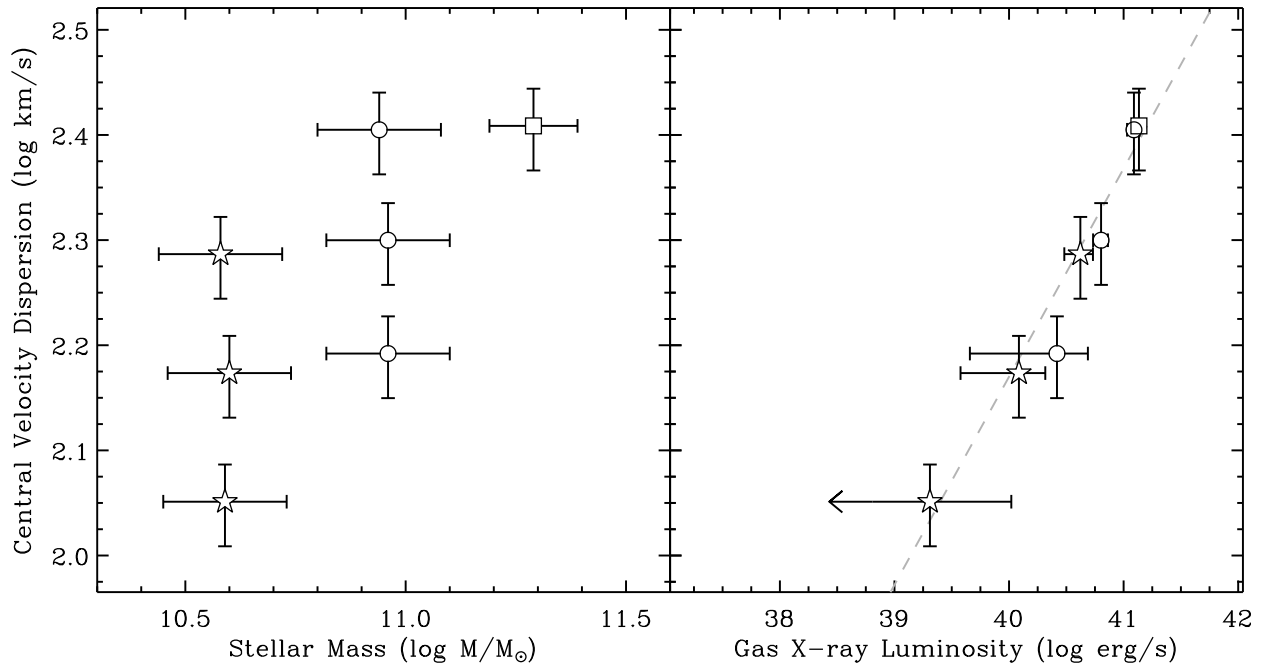}  
      \caption{Results of the stacking analysis. The left panel shows the median central stellar velocity dispersion ($\sigma_c$) as a function of the median stellar mass. The right panel depicts $\sigma_c$ as a function of the X-ray luminosity and reveals a near-perfect correlation. Note that the lowest stellar mass bin is not shown because no significant hot X-ray emitting gas is detected. The uncertainties in $\sigma_c$ represent the scatter (67th percentile) in the sample galaxies. The uncertainties in the X-ray luminosity are derived from $10^4$ jackknife simulations of the galaxies within each bin. The depicted uncertainties are computed from the 67th percentile of the simulations.}
     \label{fig:fig4}
  \end{center}
\end{figure*}

\subsection{A Tight $\sigma_{\rm c} - L_{\rm X}$ Correlation}
\label{sec:halos}
For each galaxy bin, we stacked the X-ray photons in the $0.4-2.4$ keV band RASS cutout images and their corresponding exposure maps. The stacked images are shown in Figure \ref{fig:fig3}, which reveal statistically significant detections in several bins irrespective of the median stellar mass. In particular, within the stacked images statistically significant ($4.2\sigma-7.3\sigma$) X-ray detections are found for galaxies characterized with $>190 \ \rm{km \ s^{-1}}$ median velocity dispersions. However, only moderate ($\sim2\sigma$) detections or nondetections are observed for galaxies exhibiting $<190 \ \rm{km \ s^{-1}}$ median velocity dispersions.  

To convert the net count rates to flux, we assumed an optically thin thermal plasma emission model with $kT=0.7$ keV and 0.3 solar abundance. Although this gas temperature and metal abundance are typical for elliptical galaxies \citep{humphrey06b}, we investigate whether deviations from these parameters could have any significant effect on our results. The gas temperature for galaxies in the stellar mass range of $\log M_{\rm \star}/\rm{M_{\odot}} = 10.4-11.6$ is in the range of $kT\sim0.4-0.9$ keV \citep{bogdan11}. Assuming that the lower or upper bounds of this temperature range are at fixed metal abundances would result in a $\sim3\%$ difference in the counts-to-flux conversion. Whereas most ellipticals exhibit subsolar metal abundances, some massive galaxies ($\log M_{\rm \star}/\rm{M_{\odot}} \gtrsim 11.0$) in our sample may have abundances close to the solar value. Using solar abundances instead of a $0.3$ solar value at a fixed temperature would result in a $\sim7\%$ difference in the counts-to-flux conversion. To probe the combined effects of varying temperatures and abundances, we compared two models: one with $kT=0.4$ and $0.2$ solar abundances, and another with $kT=0.9$ and solar abundance. We found that the counts-to-flux conversions for these two extreme cases are only $\sim10\%$ different. Therefore, our assumed model with $kT=0.7$ keV and 0.3 solar abundance ensures that our flux conversion is accurate to within $\sim6\%$. This value is lower than the statistical uncertainties, and hence does not influence our results in any significant way. Additionally, we note that the X-ray hardness ratios also suggest that the stacked signal is dominated by hot gas with temperatures of $kT\sim0.7-0.8$ keV (Section \ref{sec:hardness}). The average X-ray luminosities were calculated assuming the median redshift of each bin. These luminosities were corrected for the median line-of-sight column density for each bin (Table \ref{tab:list1}), which values were derived from the HI map provided by the LAB survey \citep{kalberla05}.

To derive the X-ray luminosity associated with the hot  gaseous emission, it is essential to subtract the contribution of other X-ray emitting sources. The largest contribution comes from the population of low-mass X-ray binaries (LMXBs). In elliptical galaxies, the emission from LMXBs is proportional to the stellar mass of the galaxies \citep{gilfanov04}. Therefore, we used the average LMXB luminosity function, the average LMXB spectrum \citep{irwin03}, and the stellar mass of the galaxies in each bin to derive the X-ray flux originating from unresolved LMXBs. We find that the population of unresolved LMXBs plays a minor role in the stacked signal because in those bins where hot gas is detected their contribution remains at the $3-15\%$ level. The emission from fainter compact objects, such as active binaries and cataclysmic variables, contributes at the $\lesssim 1$\% level for the elliptical galaxies in our sample \citep{bogdan08,bogdan11,sazonov06}. Hence, their contribution to the stacked signal in our sample galaxies is virtually negligible.  

To derive the uncertainties in the X-ray luminosity, we utilized Monte Carlo (MC) random simulations. For each MC simulation, we randomly stacked the X-ray images of $90\%$ of the sources present in each bin to construct the X-ray luminosity distribution of the stacked emission. From the $10^4$ simulations, we assessed the median and the 67th percentiles of the distributions to measure the uncertainties in the stacks. This process guards against the contribution from extremely luminous and underluminous sources in the sample that may dominate the observed X-ray flux. Additionally, we also derived the uncertainties using bootstrap analysis, which resulted in uncertainties identical to those obtained by the MC random simulations. 

The result of our stacking analysis is presented in Figure \ref{fig:fig4}, which reveals that, \textit{for a given stellar mass, ellipticals with the largest velocity dispersions, present the strongest X-ray luminosities}. The best-fit relation between the observed X-ray luminosity and central velocity dispersion can be described as: 
$$ \log L_{\rm X} \rm{(erg \ s^{-1})} = (33.2\pm1.6)+(3.3\pm0.7) \log \sigma_{\rm_c} \rm{(km \ s^{-1})} \ \rm{.}$$
Our X-ray stacking analysis  demonstrates a near-perfect $\sigma_{\rm c} - L_{\rm X}$ correlation with a Pearson correlation coefficient of 0.98 and $\chi^2=1.3$ for five degrees of freedom. The critical value to reject $H_0$ (i.e. no correlation between these quantities) at the $90\%$ level is $\chi^2=1.6$, and hence we conclude at the $\sim90-95\%$ confidence level that the central velocity dispersion  and the X-ray luminosity tightly correlate with each other.

\subsection{X-ray Hardness Ratios}
\label{sec:hardness}
To confirm that the detected X-ray signal originates from hot gaseous emission with sub-keV temperatures, we derive X-ray hardness ratios using the stacked RASS data. We define the hardness ratio as the ratio of count rates observed in the $1.0-2.4$ keV and $0.4-1.0$ keV energy ranges, that is $\rm{HR} = C_{\rm{1.0-2.4 keV}}/C_{\rm{0.4-1.0 keV}}$. To obtain statistically meaningful hardness ratios, we measure them only in the largest median velocity dispersion ($>190 \ \rm{km \ s^{-1}}$) bins. The hardness ratios are in the range of $\rm{HR} = 0.33-0.55$, where the largest value corresponds to more massive halos. These values are consistent with an emission spectrum dominated by hot ionized gas with $kT\sim0.6-0.8$ keV temperature, for which  $\rm{HR} = 0.35-0.5$ is expected. Moreover, observations of nearby galaxies demonstrate that these gas temperatures are typical for galaxies with stellar mass of $\sim10^{11} \ \rm{M_{\odot}}$ \citep{bogdan11}.

The derived hardness ratios are in conflict with a model in which the observed X-ray emission originates from AGNs or unresolved X-ray binaries. Indeed, a power-law model with a slope of $\Gamma=1.7$ , which is typical for AGNs or X-ray binaries, would result in $ \rm{HR} \approx 1$ at these energy ranges. We also note that an obscured AGN would have  $\rm{HR} > 1$, which is also inconsistent with our observations (see Section \ref{sec:agn} for details). Thus, the hardness ratios demonstrate that AGNs or other hard X-ray emitting components do not play a major role in the observed signal, but the observed X-ray emission is presumably dominated by hot X-ray gas with $kT\sim0.6-0.8$ keV temperature.

\subsection{Obscured AGNs are Not the Source of the \\ Stacked X-ray Signal}
\label{sec:agn}
Because X-ray stacking has been routinely used to infer the presence of heavily obscured AGNs buried at the centers of distant galaxy populations, we investigate whether the X-ray signals identified in the stacked RASS images are arising due to the presence of obscured AGNs. Therefore, we make the opposing ansatz that the observed X-ray luminosity is due to obscured BH growth. Within our stacked images, we resolve an X-ray luminosity of $\sim 3\times10^{40} \ \rm{erg \ s^{-1}}$. To convert this value to a bolometric X-ray luminosity, we assume a conversion factor $k_{\rm bol} = L_{\rm bol}/L_{\rm 0.5-2keV} = 20$ \citep{lusso12}, and a power-law model with a slope of  $\Gamma = 1.9$ and intrinsic $N_{\rm{H}} = 10^{22} \ \rm{cm^{-2}}$, which is typical of  a mildly obscured Type-2 AGN \citep[e.g.][]{risaliti99}. Thus, an observed (absorbed) X-ray luminosity of $3\times10^{40} \ \rm{erg \ s^{-1}}$ corresponds to the AGN bolometric luminosity of  $1.2\times10^{42} \ \rm{erg \ s^{-1}}$, for those galaxies with BH masses of $\sim 3\times 10^8 \ \rm{M_{\odot}}$. Such a BH mass corresponds to an Eddington limited accretion luminosity of $\sim 4\times10^{46} \ \rm{erg \ s^{-1}}$, and hence, based on the observed X-ray luminosity, a {\it lower} limit of $\sim10^{-4}$ for the Eddington ratio. Such an Eddington ratio is typical of AGNs present in star-forming spiral galaxies, which have a plentiful supply of cool gas from which to form a radiatively efficient accretion disk \citep{goulding10}. Even powerful radio AGNs that are observed in elliptical galaxies are typically found to have far lower Eddington ratios \citep{ho02}. For AGN with Eddington ratios $>10^{-4}$, we would expect to readily observe optical AGNs emission lines. However, sources such as these were removed during our galaxy selection process. 

Furthermore, our X-ray images are constructed in the soft X-ray regime, $E<2.4$ keV; such X-ray photons are particularly susceptible to even low levels of gas absorption. Hence, for an AGN to be missed at optical wavelengths, this would require relatively large levels of extinction toward the central AGN, and any soft X-ray photons arising from this AGN would still be obscured in an X-ray stack. Indeed, all previous X-ray stacking analyses have harnessed only hard $E>4 $ keV photons to infer the presence of an obscured AGN population, as the soft photons remain hidden in the stacks \citep{daddi07}. Thus, we conclude that the X-ray emission observed in the stacks does not arise because of the presence of obscured AGNs, and must instead originate from the presence of galaxy-wide hot gas.

\subsection{Robustness of the Background Subtraction} 
The nondetections of galaxies in the lowest stellar mass bin ($ \log M_{\rm{star}}/\rm{M_{\odot}}=10.0-10.4$) demonstrate that our background subtraction is robust. Indeed, most galaxies within this stellar mass bin are not expected to reside in a sufficiently massive halo that is capable of retaining significant quantities of hot X-ray gas \citep{osullivan01,mathews06,kim13}. 

To further confirm that the detected signal is not of spurious origin, we also stacked random fields within the SDSS field of view. We selected 500 random coordinates covered by the SDSS, and stacked the corresponding RASS images following our previous procedure. The obtained stacked image of random fields consists of a flat background and lacks any evidence for a statistically significant detection at the center. Hence, the detected sources, associated with the position of the stacked galaxies have a physical origin.

\begin{figure*}[!t]
  \begin{center}
\includegraphics[width=8cm]{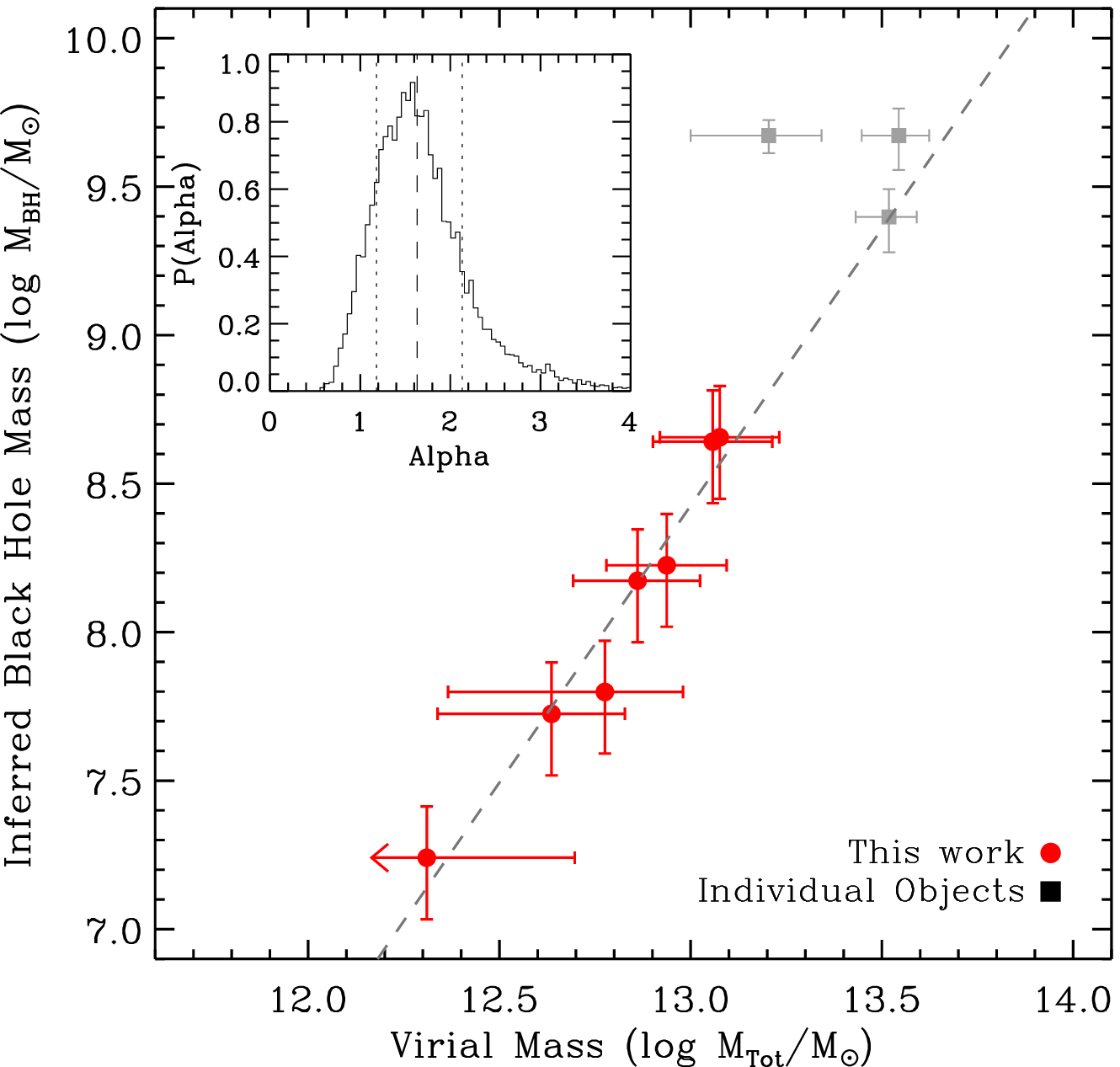}  
\hspace{1.25cm}
\includegraphics[width=8cm]{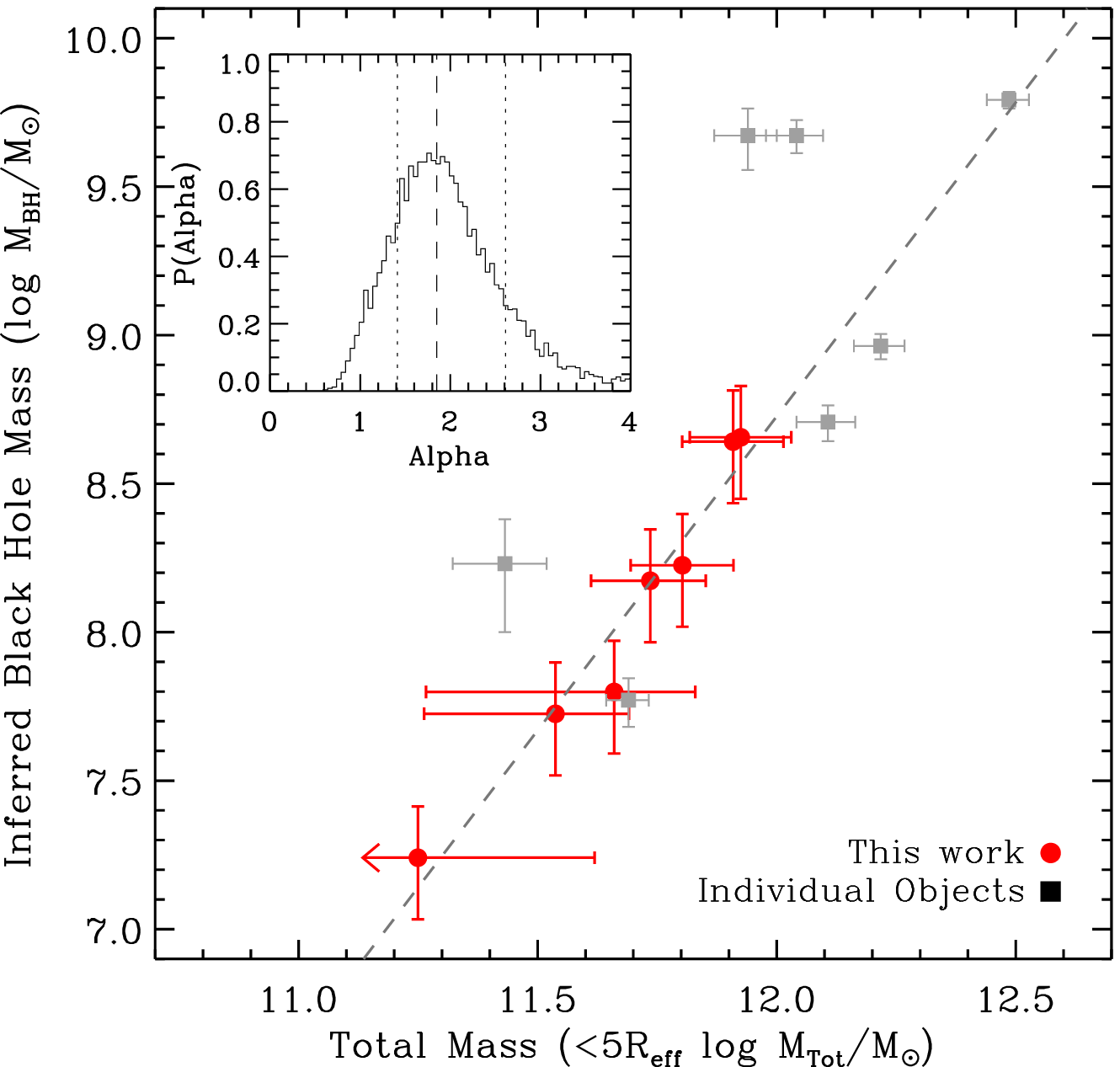}  
      \caption{Scaling relations between the inferred BH mass and the inferred total mass within the virial radius (left panel) or within $5r_{\rm eff}$ (right panel).  The best-fit $M_{\rm BH} - M_{\rm tot}$ and  $M_{\rm BH} - M_{\rm 5r_{eff}}$ relations are shown with the dashed lines. The uncertainties in total and halo mass are calculated by summing in quadrature the uncertainties measured from the MC simulations of the X-ray luminosities, shown in Figure \ref{fig:fig4}, with the rms scatter observed in the relations of \citet{mathews06} and \citet{kim13}. The uncertainties in $M_{\rm BH}$ are calculated by summing in quadrature the 67th percentile scatter in our defined velocity dispersion bins with the observed $M_{\rm BH} - \sigma_{\rm c}$ relation of \citet{gultekin09}. The insets show the probability distributions of the slopes obtained from the fitting procedure. Overplotted are individual galaxies with dynamically measured BH mass and precisely measured total halo masses.}
     \label{fig:fig6}
  \end{center}
\end{figure*}

\section{Discussion}

\subsection{Central Stellar Velocity Dispersion Correlates with the Total Halo Mass}
\label{sec:dmproxy}
X-ray observations have demonstrated that elliptical galaxies with very similar optical properties can have vastly different X-ray properties \citep{forman85,osullivan01,mathews03,bogdan11}. Specifically, these studies showed that the luminosity of the hot X-ray gas, which dominates the X-ray appearance of massive ellipticals, can vary by up to two orders of magnitude at a fixed stellar mass. However, the amount of hot X-ray gas that can be retained in the galaxy's gravitational potential well is primarily determined by the dark matter halo mass rather than by the stellar mass. Motivated by this realization, several studies explored the connection between the X-ray luminosity and halo mass of ellipticals \citep{mathews06,kim13}. These works demonstrated that higher X-ray luminosities correspond to more massive halos, and hence the X-ray luminosity of the hot gas correlates with the halo mass.

Indeed,  \citet{mathews06} established the scaling relation between the X-ray luminosity and virial mass for objects within the $\log M_{\rm{tot}}/\rm{M_{\odot}}=12.8-14.5 $ range, which sample included galaxies, galaxy groups, and poor galaxy clusters. In their study, \citet{mathews06} relied on X-ray observations of the hot gas to derive the total halo mass by assuming hydrostatic equilibrium. The obtained best-fit relation between the X-ray luminosity and the virial mass is described as $L_{\rm{X}} \propto M_{\rm{tot}}^{2.4}$. In a follow-up study, \citet{kim13} investigated a sample of  nearby early-type galaxies with a particular focus on gas-rich galaxies. In their work, \citet{kim13} utilized the results of \citet{deason12}, who determined the total mass  within $5r_{\rm{eff}}$ using optical kinematics data of globular clusters and planetary nebulae. \citet{kim13} found that for gas-rich early-type galaxies with $L_{\rm{X,gas}} > 10^{40} \ \rm{erg \ s^{-1}}$, which are relevant for our study, the $L_{\rm{X}} - M_{\rm{5r_{eff}}}$ relation is extremely tight and its rms deviation is only 0.128 dex. Their best-fit relation can be expressed as $L_{\rm{X}} \propto M_{\rm{5r_{eff}}}^{3.3}$. 

The existence of a tight $L_{\rm{X}} - M_{\rm{tot}}$ relation for gravitationally collapsed structures is expected from simple self-similar considerations \citep{kaiser86}. Because the self-similarity applies not only for the collision less dark matter particles but also for the weakly collisional hot ionized gas that is heated through shocks \citep{navarro95}, we expect to observe power-law scaling relations between the X-ray properties of the gas.  Specifically, under the assumption of hydrostatic equilibrium, the X-ray luminosity is predicted to trace the virial mass as $L_{\rm X}\propto M_{\rm{vir}}^{4/3}$. While the observed relations are steeper than that expected from analytical models, this simple model illustrates that the observed correlations are physically motivated.

Based on these considerations we conclude that, on average, the gas X-ray luminosity of our elliptical galaxy sample can be robustly used to trace the total halo mass. Combining this fact with the observed tight $\sigma_{\rm c} - L_{\rm X}$ relation (Section \ref{sec:halos}), we infer that for massive elliptical galaxies the central stellar velocity dispersion tightly correlates with the total halo mass. By definition, $\sigma_{\rm c}$ provides a dynamical estimate of the central gravitational potential of the galaxy. As such, our results suggest that the depth of the central potential well and the total gravitating mass of elliptical galaxies are intimately linked. Moreover, for elliptical galaxies, the  $\sigma_{\rm c} - L_{\rm X}$ relation is significantly tighter than the $M_{\rm{bulge}}-L_{\rm{X}}$ correlation (Section \ref{sec:secondary}). This implies that the central stellar velocity dispersion may be the fundamental galaxy property that is most tightly connected to the total halo mass. \textit{This in turn hints that for elliptical galaxies $\sigma_{\rm c}$ can more accurately trace the dark matter halos than the stellar mass.}

\subsection{BH Mass Correlates with the Total Halo Mass}
\label{sec:bhproxy}
Observational studies have repeatedly pointed out that the central stellar velocity dispersion correlates closely with the BH mass, and hence $\sigma_{\rm c}$ is a widely used surrogate for the BH mass \citep{gebhardt00,tremaine02,kormendy11}. Recent studies employing larger galaxy samples demonstrated that the $M_{\rm BH} - \sigma_{\rm c}$ scaling relation is particularly tight for elliptical galaxies \citep{gultekin09,mcconnell13}, which are relevant for our study. In these works the central velocity dispersion refers to the luminosity-weighted line-of-sight velocity dispersion measured within the half-light radius of the galaxy. The velocity dispersions provided by SDSS are measured within the $3\arcsec$ spectroscopic fiber. Our sample galaxies lie at a median distance of  $D=170$ Mpc, where the SDSS fiber size corresponds to a projected radius of $\sim1.2$ kpc. The typical half-light radius of elliptical galaxies in the studied stellar mass range is a few kpc, so the scales that are used to obtain the velocity dispersion are comparable. To confirm that the SDSS velocity dispersions are consistent with those obtained for the relatively small number of individual galaxies with dynamically measured BH mass, we cross-correlated our parent galaxy sample with the catalog of \citet{mcconnell13}. We found that three galaxies are listed in both samples. These values agree with each other within $\sim15\%$, thereby hinting that  the central stellar velocity dispersions provided by SDSS can be used to trace the BH mass. Additionally, the tight scaling relation between the central stellar velocity dispersion and the BH mass is not only supported by empirical scaling relations, but both analytical and hydrodynamical studies point out that this relation is a natural consequence of the self-regulated growth of BHs  \citep{silk98,king03,springel05,hopkins06,booth09}.

Thus, both observational and theoretical considerations confirm that the central stellar velocity dispersion can robustly trace the average BH mass for the elliptical galaxies in our study. To convert the central stellar velocity dispersion to BH mass, we used the best-fit $M_{\rm BH} - \sigma_{\rm c}$ relation obtained for elliptical galaxies \citep{gultekin09}. In Figure \ref{fig:fig6} we show the obtained scaling relation between the inferred BH and inferred total gravitating mass within the virial radius (left panel) and within five effective radii (right panel). Additionally, in Figure \ref{fig:fig6} we also overplot measurements obtained for a sample of early-type galaxies with dynamical BH mass measurements \citep{mcconnell13} and accurately measured $M_{\rm 5r_{\rm eff}}$ \citep{kim13} and $M_{\rm tot}$ \citep{humphrey06}.  

To further investigate the correlation between BH and total halo masses, we compute the slope of the best-fit relation by implementing a least-squares (LS) linear regression to the data. Because simple LS regression includes only the errors for the BH mass, we treat the halo mass for each data point as a normal distribution with width equal to the $1\sigma$ uncertainty shown in the insets of Figure \ref{fig:fig6}. We randomly select a halo mass for each point from these distributions, and repeatedly refit the data using least-squares $5\times10^4$ times. The obtained best-fit relation between the inferred BH mass and the inferred total mass within the virial radius can be described as $M_{\rm{BH}} \propto M_{\rm tot}^{1.6^{+0.6}_{-0.4}} $. The best-fit relation between the  inferred BH mass and the inferred total mass within $5r_{\rm eff}$ can be expressed as $M_{\rm{BH}} \propto M_{\rm{5r_{eff}}}^{1.8^{+0.7}_{-0.6}} $. 

These results are in good agreement with previous observational and theoretical studies. Indeed, \citet{bandara09} utilized galaxy-scale strong gravitational lenses to measure the total halo mass of galaxies and observed $M_{\rm{BH}} \propto M_{\rm tot}^{1.55\pm0.31}$. Moreover, cosmological simulations by \citet{booth10} predicted $M_{\rm{BH}} \propto M_{\rm tot}^{1.55\pm0.05}$, and the authors suggested that the halo binding energy is the fundamental property that controls the BH mass. Overall, our results are consistent with this picture. Indeed, it is feasible that the central stellar velocity dispersion acts as an intermediary to directly set the BH mass. However, as discussed in Section \ref{sec:dmproxy}, the central stellar velocity dispersion is determined by the total halo mass. Thus, in elliptical galaxies the BH mass may be (indirectly) set by the ratio of the BH and the total halo masses. As opposed to these, the scaling relation observed between the BH and the stellar mass may be the consequence of a velocity dispersion-to-halo mass \citep{vanuitert11} or the stellar-to-halo mass \citep{mandelbaum06,moster10}  relation.

\begin{figure}[!t]
  \begin{center}
\includegraphics[width=8.5cm]{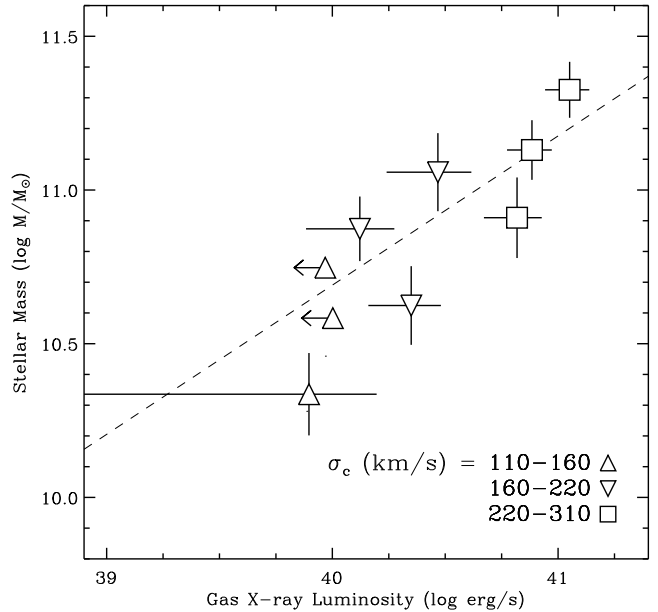}  
      \caption{Results of the stacking analysis showing the stellar mass as a function of the X-ray luminosity for bins with fixed central velocity dispersion. The applied velocity dispersion ranges are marked with different symbols. The $M_{\rm{bulge}}-L_{\rm{X}}$ relation reveals a weak correlation with a Pearson correlation coefficient of 0.73. Note that galaxies with similar central velocity dispersions have very similar X-ray luminosities despite their significantly different stellar masses. Along with the results of Figure \ref{fig:fig6}, this implies that the primary scaling relation is that between the BH mass and the total halo mass, and the stellar bulge plays a secondary role.}
     \label{fig:fig5}
  \end{center}
\end{figure}

\subsection{A Weaker Correlation between the BH and Stellar Masses}
\label{sec:secondary}
To investigate the importance of the stellar mass in producing the observed X-ray luminosity, we repeat our stacking analysis by  defining the galaxy bins in an inverted manner. Specifically, we divide the galaxy sample into four equal-width central velocity dispersion bins in the $110-310 \ \rm{km \ s^{-1}}$ range. Similarly to our main analysis, these bins are further split based on their stellar mass, and the X-ray photons were stacked along with the corresponding exposure maps. Following our earlier procedure, we extracted the source counts using a circular region with four-pixel radius, and obtained the background level from a circular annulus with $8-14$ pixel radii. For bins with X-ray luminosity of $\log L_{\rm{X}}/(\rm{erg \ s^{-1}})>40.1$ statistically significant ($3.2\sigma-6.0\sigma$) detections are observed, and for lower X-ray luminosities, moderately significant detections or nondetections are observed. The conversion of observed net counts to flux and the subtraction of LMXBs were also performed as before (see Section \ref{sec:halos}).

Figure \ref{fig:fig5} shows the resulting $M_{\rm{bulge}}-L_{\rm{X}}$ scaling relation, which reveals a significantly weaker correlation than that obtained between the central velocity dispersion and the X-ray luminosity. Indeed, for a given stellar mass, the X-ray luminosity exhibits large variations: for example, at $\log M_{\rm \star}/\rm{M_{\odot}} \sim 10.9$ we observe an order of magnitude difference in X-ray luminosity. We stress that this result refers to average galaxy populations because our stacking analysis, by definition, removes the distorting effects of individual outlier galaxies. The statistical analysis of the $M_{\rm{bulge}}-L_{\rm{X}}$ relation produces a Pearson correlation coefficient of 0.73. For five degrees of freedom, the critical value to rule out $H_0$ (i.e. no correlation between these quantities) at only the $90\%$ confidence level would require $H_0>0.81$. Additionally, the best-fit relation has a reduced $\chi^2=2.2$ for five degrees of freedom, which also hints at the large scatter. These results are in accord with earlier studies, which showed that for individual galaxies, at a given stellar mass, the luminosity of the hot X-ray gas may vary as much as two orders of magnitude  \citep{forman85,osullivan01,mathews03,bogdan11}. Beyond the large scatter in the $M_{\rm{bulge}}-L_{\rm{X}}$ relation, Figure \ref{fig:fig5} also reveals that galaxies with similar mean central stellar velocity dispersions have very similar X-ray luminosities despite their markedly different stellar masses. This result is in excellent agreement with the conclusions presented in Figure \ref{fig:fig4}, and illustrates that the stellar mass may play a more secondary role in determining the BH mass.  

In the present work we cannot probe whether a fundamental plane exists that folds in the stellar mass of the elliptical galaxies. In this picture, the physical relation is dominated by the connection between the central potential and the dark matter, but also with the stellar mass playing a further, possibly secondary, role \citep{cappellari13}. However, given the shallow RASS observations, the connection between the inferred BH mass and total halo mass could only be studied in a relatively small number of bins. Because of this limitation, the further study of the fundamental plane is beyond the scope of the current paper. However, in the near future, the sensitive all-sky survey performed by \textit{eROSITA} will allow us to split our galaxy sample into significantly smaller bins, and hence explore the parameter space in greater detail.

\section{Conclusions}
In this work we utilized a sample of 3130 elliptical galaxies observed by the SDSS and the RASS, and probed whether BHs exhibit a correlation with their dark matter halos. Our results can be summarized as follows: 

\begin{enumerate}
\item We grouped our galaxy sample based on stellar mass and central velocity dispersion, and found that the stacked X-ray images exhibit statistically significant detections for bins with $\sigma_{\rm c}>190 \ \rm{km \ s^{-1}}$ and moderate detections or nondetections for bins with $\sigma_{\rm c}<190 \ \rm{km \ s^{-1}}$, irrespective of the median stellar mass of the galaxies in the bins. Based on the SDSS and stacked RASS data, we established that for massive elliptical galaxies the central stellar velocity dispersion exhibits a significantly tighter correlation with the X-ray luminosity of the hot gas than with the stellar mass.
\item We used the X-ray luminosity as a surrogate of the total halo mass, and concluded that it may be the central stellar velocity dispersion, and hence the central gravitational potential, that is most closely related to the dark matter halos of elliptical galaxies. This hints that for elliptical galaxies the central stellar velocity dispersion may trace the large-scale dark matter halo more accurately than the stellar mass.
\item By utilizing the central stellar velocity dispersion as a tracer of the BH mass, we established that the inferred BH and total halo masses tightly correlate with each other. The best-fit slopes between the inferred BH mass and the total mass within the virial radius or $5r_{\rm eff}$ can be described as  $M_{\rm{BH}} \propto M_{\rm tot}^{1.6^{+0.6}_{-0.4}} $ or $M_{\rm{BH}} \propto M_{\rm{5r_{eff}}}^{1.8^{+0.7}_{-0.6}}$, respectively. These results hint that in elliptical galaxies the central stellar velocity dispersion may act as an intermediary and directly set the BH mass. Because the central stellar velocity dispersion is determined by the dark matter halo mass, the BH mass may be (indirectly) set by dark matter halo mass.
\end{enumerate}

\bigskip
\begin{small}
\noindent
Acknowledgements. The authors thank the anonymous referee for the excellent and insightful suggestions that have greatly improved the paper and its discussions. The authors thank W. Forman, J. Greene, and R. Hickox for helpful discussions. This research has made use of data, software and Web tools obtained from the High Energy Astrophysics Science Archive Research Center (HEASARC), a service of the Astrophysics Science Division at NASA/GSFC and of the Smithsonian Astrophysical Observatory's High Energy Astrophysics Division. This work makes use of data from the SDSS-III survey. Funding for SDSS-III has been provided by the Alfred P. Sloan Foundation, the participating institutions, the National Science Foundation, and the U.S. Department of Energy Office of Science. The SDSS-III Web site is http://www.sdss3.org/. \'AB acknowledges support provided by NASA through Einstein Postdoctoral Fellowship grant number PF1-120081 awarded by the CXC, which is operated by the Smithsonian Astrophysical Observatory for NASA under contract NAS8-03060. ADG acknowledges funding from NASA grant AR3-14016X.  \end{small}

\end{document}